\documentclass[doublecol]{epl2}

\usepackage{amssymb}
\usepackage{graphicx}
\usepackage{dcolumn}
\usepackage{bm}
\usepackage{amsmath}

\title{Magnetic impurities in the two-band $s_\pm$-wave superconductors}
\author{Jian Li and Yupeng Wang}
\institute{Beijing National Laboratory for Condensed Matter Physics,
and Institute of Physics, Chinese Academy of Sciences, Beijing
100190, China}

\pacs{74.20.-z}{Theories and models of superconducting state}
\pacs{74.62.Dh}{Effects of crystal defects, doping and substitution}

\abstract {We investigate the effects of magnetic impurities in a
superconducting state with $s_\pm$ pairing symmetry. Within a
two-band model, we find that the intra-band magnetic scattering
serves as a pair breaker while the inter-band magnetic scattering
preserves pairing and hardly affects transition temperature in the
Born limit. We also show that the same physics can persist beyond
the weak scattering region. Our results coincide with recent
experimental measurements in iron-based superconductors and thus
provides an indirect evidence of the possible $s_\pm$ pairing
symmetry in these materials.}

\begin{document}
 \maketitle

The iron-based superconductors have attracted much attention since the
compound $LaFeAsO_{1-x}F_{x}$ was found superconducting with $T_c=26K$\cite%
{hosono}. Accompanied with the increasing transition temperature over 40K%
\cite{genfu} and finally up to 55K\cite{zhao2}, these new superconductors
are considered as the second family of high-temperature superconductors
after the cuprates. Though much efforts have been made and great progress
has been achieved experimentally, the superconducting mechanism and the
pairing symmetry are still unclear because of the complexity of
multi-orbital nature and possible strong electron-electron correlation in
these materials. The angle-resolved photoemission spectroscopy (ARPES)\cite%
{ding}, the Andreev reflection\cite{chien} and the penetration depth\cite%
{Hashimoto} experiments directly showed fully gaped superconductivity while
the nuclear-magnetic-resonance (NMR) indicates a strong deviation from that
of single band $s$-wave superconductors\cite{NMR}. Simultaneously, the $%
s_\pm $ pairing symmetry was proposed and widely discussed by many
authors theoretically\cite{Mazin}. With this kind of pairing
symmetry, fully opened gap located in the hole pocket around the
$\Gamma $ point and electron pocket around the $M$ point have
opposite signs. The $s_\pm $ pairing symmetry is consistent with the
ARPES results and also coincides with the NMR results providing the
presence of strong impurities\cite{Parker} and thus becomes a very
promising candidate to account for the main physics of the
iron-based superconductors.

To distinguish the pairing symmetry experimentally, we note that the effects
of impurities in superconducting states can be very different for different
pairing symmetries. Very recently, superconductivity in both 1111 and 122
systems has been induced by doping magnetic elements $Co$\cite{Cwang,sefat2}%
. These experiments share some common features. First, the superconductivity
in these materials shows high tolerance with the disorder induced by the $Co$%
-doping. Second, the suppression of the transition temperature is
not so significant but still much stronger than the $F$-doped
materials. The second point can be easily understood because the
$F$-doping happens in the $LaO$ layers while the $Co$-doping happens
in the $FeAs$ layers which are crucial to the superconductivity.
However, a later experiment\cite{Xu} on $Zn$-doped $LaFeAsO$
shows that the superconductivity is almost unperturbed by the $Zn$%
-doping, though it happened in the $FeAs$ layers. This strongly indicates
that the non-magnetic impurities are unlikely to affect the
superconductivity, in accordance with s-wave pairing symmetry. Meanwhile,
magnetic impurities can depress the transition temperature but not so
significant as in the conventional s-wave superconductors. A. E. Karkin et
al used fast neutron irradiation to induce disorder in $LaFeAsO_{0.9}F_{0.1}$
and found the depression of transition temperature can be qualitatively
described by Abrikosov-Gorkov (AG) theory with magnetic impurities\cite%
{Karkin}. However, such depression is also much slower than that predicted
by the AG theory (see Fig.10 in \cite{Karkin}).

To solve this puzzle, we propose a model in this letter to describe the
behavior of the magnetic impurities in an $s_\pm$-wave superconductor and
show how the inter-band magnetic impurity scattering processes may preserve
the pairs and effectively weaken the reduction of transition temperature.
Our theoretical results give a reasonable explanation of the experimental
data.

We start with a model consisting of two perfect nested bands, i.e.,
an electron Fermi pocket and a hole Fermi pocket. The details of the
band structure is neglected and the superconducting order parameters
located in each band have same magnitude but reversed signs as in
Refs.\cite{Han,senga}. For convenience, we introduce the Nambu
vector
\begin{equation*}
\psi _{\mathbf{k}}=(c_{\mathbf{k}\uparrow },c_{\mathbf{k}\downarrow },d_{%
\mathbf{k}\uparrow },d_{\mathbf{k}\downarrow },c_{-\mathbf{k}\uparrow
}^{\dag },c_{-\mathbf{k}\downarrow }^{\dag },d_{-\mathbf{k}\uparrow }^{\dag
},d_{-\mathbf{k}\downarrow }^{\dag }),
\end{equation*}%
here $c_{\mathbf{k}\sigma }$ and $d_{\mathbf{k}\sigma }$ are the
annihilation operators in the electron band and the hole band and the band
energy $\epsilon _{c,\mathbf{k}}=-\epsilon _{d,\mathbf{k}}=\epsilon _{%
\mathbf{k}}$, respectively. Here we treat the magnetic impurities as
localized spins in the classical limit and the quantum (Kondo)
effect of impurities is not under our consideration. In this limit,
the magnetic impurity is equivalent to the local magnetic field.
Then the interaction matrix due to magnetic impurity scattering is
assumed to be
\begin{equation*}
V=\frac{1}{2}\left(
\begin{array}{cccc}
J_{1}\mathbf{\sigma }\cdot \mathbf{S} & J_{2}\mathbf{\sigma }\cdot \mathbf{S}
& 0 & 0 \\
J_{2}\mathbf{\sigma }\cdot \mathbf{S} & J_{1}\mathbf{\sigma }\cdot \mathbf{S}
& 0 & 0 \\
0 & 0 & J_{1}\sigma _{2}\mathbf{\sigma }\cdot \mathbf{S}\sigma _{2} &
J_{2}\sigma _{2}\mathbf{\sigma }\cdot \mathbf{S}\sigma _{2} \\
0 & 0 & J_{2}\sigma _{2}\mathbf{\sigma }\cdot \mathbf{S}\sigma _{2} &
J_{1}\sigma _{2}\mathbf{\sigma }\cdot \mathbf{S}\sigma _{2}%
\end{array}%
\right)
\end{equation*}%
Here $\mathbf{\sigma }$ and $\mathbf{S}$ denotes the spin operator of the
electrons and magnetic impurities, respectively. Both the intra-band ($J_{1}$%
) and the inter-band ($J_{2}$) exchange coupling constants are
positive and isotropic in our consideration.

Following the AG theory\cite{AG}, the renormalized two-band BCS Green's
function with randomly distributed impurities of finite concentration $%
n_{imp}$ reads:
\begin{eqnarray}
&&G^{-1}(\mathbf{k},\omega )=G_{0}^{-1}(\mathbf{k},\omega )-\sum
(\omega )
\notag \\
&=&\left(
\begin{array}{cccc}
i\widetilde{\omega }-\epsilon _{\mathbf{k}} & x & -i\widetilde{\Delta }%
\sigma _{2} & 0 \\
x & i\widetilde{\omega }+\epsilon _{\mathbf{k}} & 0 & i\widetilde{\Delta }%
\sigma _{2} \\
i\widetilde{\Delta }\sigma _{2} & 0 & i\widetilde{\omega }+\epsilon _{%
\mathbf{k}} & x \\
0 & -i\widetilde{\Delta }\sigma _{2} & x & i\widetilde{\omega }-\epsilon _{%
\mathbf{k}}%
\end{array}%
\right) ,
\end{eqnarray}%
where $G_{0}(\mathbf{k},\omega )$ is the Green's function without the
impurity; $\sum (\omega )$ is the self energy; $\widetilde{\omega }$ and $%
\widetilde{\Delta }$ is the renormalized frequency and
superconducting order parameter, respectively. The parameter $x$ is
the inter-band scattering induced contribution to the self-energy
which can be determined self-consistently. In the Born approximation
the self-energy can be written as\cite{Aff}
\begin{eqnarray*}
&&\sum (\omega )=n_{imp}\int \frac{d^{2}k^{\prime }}{(2\pi )^{2}}%
\left\langle VG(k^{\prime },\omega )V\right\rangle _{I}
\end{eqnarray*}
where $\left\langle ...\right\rangle _{I}$ means averaging the impurity
position. Now the renormalized Green's function can be given
self-consistently with
\begin{eqnarray}
\widetilde{\omega } &=&\omega +(\frac{ix}{2\tau _{3}}+\frac{\widetilde{%
\omega }}{2\tau _{1}})\frac{1}{\sqrt{\widetilde{\Delta }^{2}+\widetilde{%
\omega }^{2}+x^{2}}}, \\
\widetilde{\Delta } &=&\Delta -\frac{1}{2\tau _{2}}\frac{\widetilde{\Delta }%
}{\sqrt{\widetilde{\Delta }^{2}+\widetilde{\omega }^{2}+x^{2}}}, \\
x &=&(\frac{i\widetilde{\omega }}{2\tau _{3}}-\frac{x}{2\tau _{1}})\frac{1}{%
\sqrt{\widetilde{\Delta }^{2}+\widetilde{\omega }^{2}+x^{2}}},
\end{eqnarray}%
and the spin-flip scattering times:
\begin{eqnarray*}
\frac{1}{\tau _{1}} &=&\frac{(J_{1}^{2}+J_{2}^{2})}{2}\pi N_{F}n_{imp}S(S+1),
\\
\frac{1}{\tau _{2}} &=&\frac{(J_{1}^{2}-J_{2}^{2})}{2}\pi N_{F}n_{imp}S(S+1),
\\
\frac{1}{\tau _{3}} &=&J_{1}J_{2}\pi N_{F}n_{imp}S(S+1),
\end{eqnarray*}%
with $N_{F}$ the density of states at the Fermi surface.

With the above equations we obtain
\begin{equation*}
\frac{\omega }{\Delta }=u[1-\frac{1}{\tau _{S}\Delta }\frac{1}{\sqrt{%
1+u^{2}-(\frac{I}{\widetilde{\Delta }})^{2}}}],
\end{equation*}%
with $u=\frac{\widetilde{\omega }}{\widetilde{\Delta }}$ and $I=ix$. The
effective pair-breaking parameter is $\alpha\equiv\frac{1}{\tau _{S}\Delta }%
=(\frac{I}{ \widetilde{\omega }}\frac{1}{2\tau _{3}}+\frac{1}{2\tau _{1}}+%
\frac{1}{2\tau _{2}})\frac{1}{\Delta }$.

Now it is clear that the Green's function contains two terms: one is the
conventional term as in the single-band $s$-wave superconductors with the
renormalized frequency and superconducting order parameter and the other one
is the contribution induced by the inter-band magnetic impurity scattering
processes. If the inter-band scattering term $x$ in the renormalized Green's
function is neglected, the effective pair-breaking parameter reads $%
\alpha=\alpha _{0}=(\frac{1}{2\tau _{1}}+\frac{1}{2\tau _{2}})\frac{1}{
\Delta }\propto J_{1}^{2}$, quite similar to that in the conventional AG
theory. However, a non-zero $J_2$ will change the situation significantly as
we shall show below.

Making $i\omega \rightarrow \omega $, the density of states (DOS) is given
by:
\begin{eqnarray*}
N(\omega ) &=&-\frac{1}{\pi }Im\int \frac{d^{2}k}{(2\pi )^{2}}%
G_{11}^{R}(k,\omega ) \\
&=&N_{F}Im(\frac{\widetilde{\omega }}{\sqrt{\widetilde{\Delta }^{2}-%
\widetilde{\omega }^{2}+x^{2}}}).
\end{eqnarray*}%
Numerical results of the DOS for given $\alpha _{0}$ and $\lambda
=J_{2}/J_{1}$ are shown in Fig.1.
\begin{figure}[tbh]
\label{Fig.1}\includegraphics[width=3.4in] {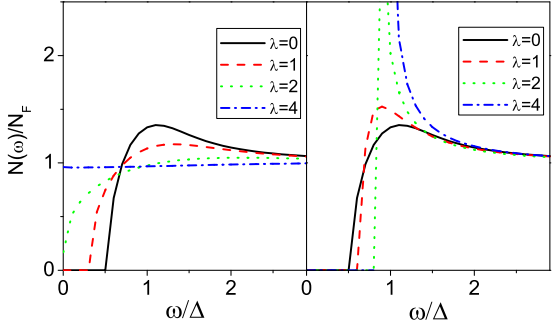}
\caption{(Color online) The density of states for $\protect\alpha
_{0}=0.2$ and $\protect\lambda =J_{2}/J_{1}$. The left part is for
the $s$-wave pairing symmetry and the right one is for $s_{\pm
}$-wave pairing symmetry.}
\end{figure}
The densities of states for both the conventional two-band $s$-wave case and
the $s_{\pm }$-wave case are calculated. The numerical results clearly show
that in the conventional $s$-wave case (left part of Fig.1), the inter-band
scattering also depresses the superconducting gap. However, accompanied with
the increasing of $J_{2}$, the superconducting gap is growing larger in $%
s_{\pm }$-wave case. This strongly indicates that $J_{1}$ and
$J_{2}$ have opposite effects on the gap of $s_{\pm }$-wave
superconductors. It seems that the inter-band magnetic impurity
scattering played as a pair repairer in our system. To make that
clearer we calculated the superconducting gap in finite temperature.
Here the superconducting gap $\Delta $ in equation (3) should be
replaced by $\Delta (T)$, and $\Delta (T) $ is determined by
\begin{equation}
\Delta (T)=V^{SC}N_{F}\pi T\underset{m}{\sum }\frac{\widetilde{\Delta }}{%
\sqrt{\widetilde{\Delta }^{2}+\widetilde{\omega }_{m}^{2}+x^{2}}}.
\end{equation}%
where $V^{SC}$ is the coupling constant and $\omega _{m}=(2m+1)\pi
T$.

Solution of (2)-(5) gives the finite temperature gap $\Delta (T)$.
The intra- and inter-band magnetic impurity effect to $\Delta (T)$
are shown in Fig.2. In Fig.2(a) only intra-band impurity scattering
exists and $\frac{1}{\tau _{1}}=\frac{1}{\tau _{2}}\propto
J_{1}^{2}$. When we increase the intra-band scattering, the finite
temperature gap and the transition temperature become smaller. This
case is in accordance with conventional $s$-wave
superconductors\cite{Sk}. In Fig.2(b) things will be reversed if we
settle down the
intra-band scattering and increase the inter-band one. $\Delta (T)$ and $%
T_{C}$ becomes larger with $J_{2}$ increasing.

\begin{figure}[tbh]
\label{Fig.2}\includegraphics[width=3.0in] {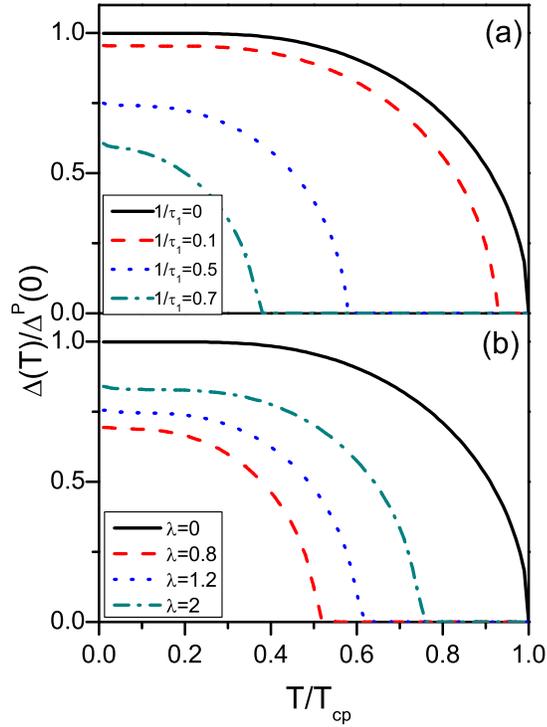}
\caption{(Color online) The superconducting gap in finite
temperature with different intra- or inter-band impurity scattering
amplitudes. The solid black line in both (a) and (b) is the two
bands BCS case without impurities. $\Delta^P(0)$ is the
superconducting gap without impurities in zero temperature.
(a)$J_{2}=0$. (b)  The solid black line: $n_{imp}=J_1=0$; others:
$\pi n_{imp}N_FS(S+1)=0.7$, $J_{1}=1$.}
\end{figure}

Such an effect also reflects in the transition temperature
\begin{equation*}
ln\frac{T_{c}}{T_{cp}}=\sum_{m=0}[\frac{1}{(m+\frac{1}{2})\sqrt{1-(\frac{I}{%
\widetilde{\omega }_{m}})^{2}}+\frac{1}{\tau _{S}}\frac{1}{2\pi T_{c}}}-%
\frac{1}{m+\frac{1}{2}}],
\end{equation*}%
where $T_{c}$ and $T_{cp}$ are the transition temperatures with and without
magnetic impurities, respectively. The numerical results of the transition
temperature vs. $\alpha _{0}$ is depicted Fig.3.
\begin{figure}[tbh]
\label{Fig.3}\includegraphics[width=3.1in] {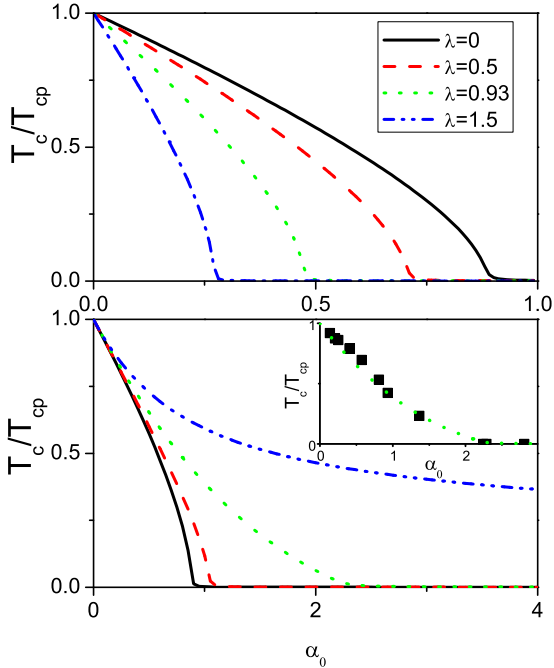}
\caption{(Color online) The reduced transition temperature for the
conventional $s$-wave (above) and $s_{\pm }$-wave (below). When $\protect%
\lambda =0$ the conventional AG behavior is regained. Inset: fitting
of the experimental data (filled squares) from Ref. [13] with
$\protect\lambda =0.93$}
\end{figure}
Once the $\lambda =J_{2}/J_{1}$ increasing, the depression of the transition
temperature is effectively speeded up in the $s$-wave superconductor but
slowed in the $s_{\pm }$-wave superconductor. In experiments, $Co$-doping
introduces extra carriers and modifies the crystal structure\cite{Cwang}. As
those changes may be crucial to the superconductivity, it is quite difficult
to examine whether other issues may play a role in the robustness of
superconductivity. However, a very recent neutron irradiation experiment on $%
LaO_{0.9}F_{0.1}FeAs$ provides a chance to check our theory since in
this experiment both the crystal structure and carrier density of
the sample are almost unchanged. The comparison is shown in Fig.3.
The experimental data is fitted quite well with our theory
quantitatively. In the real iron-based materials, with the
increasing of the $Co$-doping, the hole pocket becomes smaller and
the electron pocket becomes larger because of the shift of the
chemical potential. This will weaken the inter-band magnetic
scattering and preserving of $T_{c}$.

If $\lambda >1$, $T_{c}$ goes down to a finite value rather than zero when $%
\alpha _{0}$ is very large. This indicates that in the Born approximation
the \emph{superconductivity can never be destroyed by magnetic impurities if
the inter-band scattering is stronger than the intra-band one in the $s_{\pm
}$ -wave superconductors}. To clarify this issue further, we consider a
single classical spin in an $s_{\pm }$-wave superconductor, which can be
treated with more accuracy. The T-matrix in this case is given by
\begin{equation*}
T=V[1-G^{0}(0,\omega )V]^{-1}.
\end{equation*}%
The energy of the bound state induced by the single magnetic impurity in an $%
s_{\pm }$ superconductor is determined by the pole of the T-matrix
\begin{equation}
\frac{\omega _{0}}{\Delta }=\pm \frac{1-\alpha _{1}+\alpha _{2}}{\sqrt{%
1+2\alpha _{1}+\alpha _{1}^{2}-2\alpha _{1}\alpha _{2}+2\alpha _{2}+\alpha
_{2}^{2}}},
\end{equation}%
while the position of the bound state for the conventional $s$-wave
superconductor is $\frac{\omega _{0}}{\Delta }=\pm \frac{1-\alpha _{3}}{%
1+\alpha _{3}}$ with
\begin{eqnarray*}
\alpha _{1} &=&(\frac{\pi J_{1}SN_{F}}{2})^{2}, \\
\alpha _{2} &=&(\frac{\pi J_{2}SN_{F}}{2})^{2}, \\
\alpha _{3} &=&[\frac{\pi (J_{1}\pm J_{2})SN_{F}}{2}]^{2}.
\end{eqnarray*}%
If $J_{2}=0$, i.e., only the intra-band impurity scattering exists\cite%
{jiangping}, the bound state energy (BSE) is $\frac{\omega _{0}}{\Delta }%
=\pm \frac{1-\alpha _{1}}{1+\alpha _{1}}$, which falls into the gap\cite{CS}
and we recover the Yu-Shiba-Rusinov solution\cite{Yu,Shiba,Rusinov}. If only
the inter-band impurity scattering exists, The BSE is $\frac{\omega _{0}}{%
\Delta }\rightarrow \pm 1$, i.e., locates at the gap edge. Generally, the
BSE falls into the gap. Note for the conventional two-band $s$-wave
superconductors, the bound state splits into two branches in each band due
to the inter-band magnetic scattering. However, for the $s_{\pm }$ case,
there is only one bound state and increasing $J_{1}$ pushes the BSE to the
Fermi energy side while increasing $J_{2}$ pushes the BSE to the gap edge
side. With a finite impurity concentration, an impurity band\cite{Shiba}
will be expanded around the position of the BSE. The superconducting gap can
be suppressed (enlarged) by increasing $J_{1}$ ($J_{2}$). This analysis
supports our Born approximation result.
\begin{figure}[tbh]
\label{fig4}\includegraphics[width=3.4in] {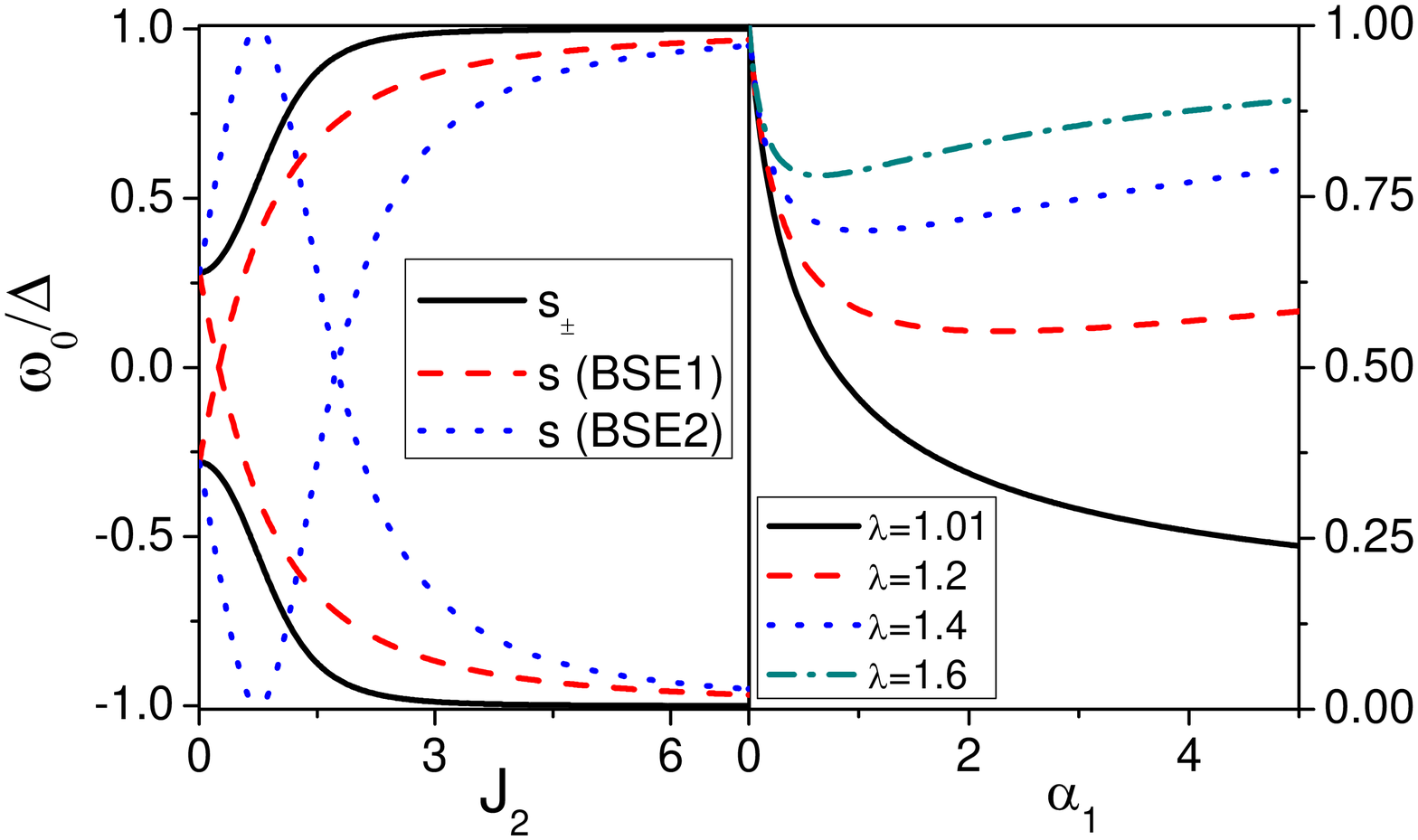} \caption{(Color
online) Left: The position of the bound state energy vs. the
inter-band coupling constant $J_{2}$ in conventional $s$ and $s_{\pm
}$ superconductors. The two branches of the bound states in $s$ case
are labeled as BSE1 and BSE2 and $J_{1}=0.75$. Right: Change of the
position of the bound state energy with $\protect\alpha _{1}$ and
$\protect\lambda =J_{2}/J_{1}$}
\end{figure}

When $\lambda >1$, from Eq.(3) we can see that
\begin{eqnarray*}
&&\frac{1-\alpha _{1}+\alpha _{2}}{\sqrt{1+2\alpha_1 +\alpha
_{1}^{2}-2\alpha _{1}\alpha _{2}+2\alpha _{2}+\alpha _{2}^{2}}} \\
&=&\frac{1}{\sqrt{1+\frac{4\alpha _{1}}{[1+(\lambda ^{2}-1)\alpha _{1}]^{2}}}%
}\geq \sqrt{1-\lambda ^{-2}},
\end{eqnarray*}
which means in this case the BSE has always a positive value and can never
reach the Fermi level. Such fact preserves the finite superconducting gap
even with a finite impurity concentration of impurities and explains why $%
T_{c}$ goes to a finite value rather than zero with the increasing of $%
\alpha _{0}$ in Fig.3. The position of the BSE with increasing
$\alpha_{1}$ and different $\lambda$ is shown in the right part of
Fig.4. Considering the impurity band expanded around the position of
BSE\cite{Shiba}, the region where always exists a finite gap should
be in $\lambda \geq 1$. The discussion on the single impurity
problem makes our conclusion to a broader region beyond the weak
scattering limit.

It is well known that the magnetic impurity is a pair-breaker in the
conventional spin singlet superconductors because it breaks the
time-reversal symmetry and nonmagnetic impurity can break the pairs in $d$%
-wave superconductors because it smears the anisotropy of the order
parameter. If we neglect $J_1$ and look back to the effect of the inter-band
impurity scattering, we find that both of the time-reversal breaking and
sign reversal of the order parameters exist but canceled each other in the $%
s_\pm$-wave superconductors. In fact, due to the sign reversal of the order
parameters, the wave functions in different pockets have a spin up-down
exchange. Therefore, the inter-band spin-flip scattering process caused by
magnetic impurity is the same as the intra-band nonmagnetic scattering
process which preserves the spins and does not break the pairs. This is in
accordance with the "reversed AG theory" in the inter-band channel which is
proposed and noticed before\cite{Mazin2}. Though the inter-band magnetic
scattering has no effect on pair breaking but $J_1$ can still suppress the
superconducting gap and $T_c$. In our system which is based on one electron
band and one hole band with the $s_\pm$-wave pairing symmetry, the emerge of
the inter-band scattering via impurities is inevitable and makes the
inter-band magnetic scattering behaves like a "pair repairer" and weakens
the depression of the superconductivity.

Based on our theoretical prediction, one may distinguish the pairing
symmetry in experiments by substituting magnetic ions out of but coupled to
the $FeAs$ layers and the $T_c$ depression must behaves quite differently in
$s$-wave and $s_\pm$-wave superconductors. A very recent experiment\cite{PC}
showed that $BaFe_{1-x}Co_xAs_2$ is always more robust and have larger
superconducting region than $BaFe_{1-x}Ni_xAs_2$, though $Ni $ doping can
result in smaller $c$-axis and bring more extra electrons into the
superconducting layer. In another recent experiment\cite{YM}, the compound $%
Ba_{1-x}K_{x}Fe_2As_2$ shows a very large superconducting area with doping
(from x=0.1 to x=1.0). These experiments also indicate the importance of the
inter-band magnetic scattering. However, the in-plane doped magnetic
impurities may affect other factors relevant to the superconductivity which
make the situation unclear. One can also distinguish the $s_\pm$ and
conventional $s$ pairing symmetry in two band superconductors by detecting
the bound state energy. There is one bound state in $s_\pm$ case but two in $%
s$ case. Besides, with the increasing of $J_{2} $, the BSE will move to the
gap edge side in $s_\pm$ case. In conventional $s$ case, BSE1 will move to
the the Fermi edge side and quickly a quantum phase transition happens\cite%
{Balatsky}, but BSE2 first moves to the gap edge when $J_2<J_1$ and
then goes to the Fermi energy side when $J_2>J_1$. Such behavior is
shown in the left part of Fig.4. The position of BSE near the
magnetic impurity can be detected from the tunneling spectra with a
low-temperature scanning tunneling microscope (STM) which has been
used to detect the Yu-Shiba-Rusinov bound state
successfully\cite{ali}.

In summary, the magnetic impurity effect in the two-band
superconductors with a perfect nesting effect is studied for both
the conventional $s$-wave pairing and the $s_\pm$-wave pairing.
Under the condition of the same magnitude of order parameters in
each band, it is found that the depression behaviors to the
superconductivity with these two different pairing symmetries are
quite different due to the existence of the inter-band impurity
scattering, which is almost inevitable in the multi-band systems.
Our theory surprisingly coincides with the neutron irradiation
experiment quantitatively and we believe it can also explain why the
iron-based superconductors show high tolerance with magnetic
impurities in many recent experiments. This provides an indirect
method to detect the pairing symmetry of the $FeAs$ superconductors
in experiments.

We would like to thank L. Yu, J. P. Hu, C. Fang, W. Tsai, X. L. Cui,
Q. F. Sun and T. Xiang for fruitful discussions and suggestions. We
also thank A. E. Karkin et al for providing their experimental data
before publication. This work was financially supported by NSFC, CAS
and 973-project of MOST of China.

\end{document}